\documentclass[amstex,12pt,epsfig,latexsym]{article}
\usepackage{amsmath}
\usepackage{epsfig}

\makeatletter
\newcommand{\gsim}{\raise.3ex\hbox{$>$\kern-.75em\lower1ex\hbox{$\sim$}}}
\newcommand{\lsim}{\raise.3ex\hbox{$<$\kern-.75em\lower1ex\hbox{$\sim$}}}

\begin{document}

\title{Hidden sector effects on double Higgs production near
threshold at the LHC}

\author{A.~C.~A.~Oliveira\footnote{\texttt{xanda@ift.unesp.br}} and
R.~Rosenfeld\footnote{\texttt{rosenfel@ift.unesp.br}} }

\maketitle

\begin{center}
\vskip 3pt  \emph{Instituto de F\'{\i}sica Te\'orica \\
Universidade Estadual Paulista\\
Rua Dr. Bento T. Ferraz, 271, 01140-070,
S\~ao Paulo, Brazil} 
\end{center}

\vskip 3pt
\begin{abstract}
In this letter we study a novel effect of a hidden sector coupling 
to the standard model Higgs boson: an enhancement of the Higgs pair production
cross section near threshold due to bound state effects. 
After summing the ladder contributions of the hidden sector to the effective $ggHH$ coupling, we find
the amplitude for gluon-gluon scattering via a Higgs loop. We relate this amplitude
to the double Higgs production cross section via the optical theorem. 
We find that enhancements of the ${\cal O}(100)$ for the partonic cross section near the threshold region can be 
obtained for a hidden sector
strongly coupled to the Higgs boson. The corresponding cross section at the LHC can be as large as
${\cal O} (10)$ times the SM result for extreme values of the coupling.
The detection of such an effect could in principle lead to important information about the hidden sector.
\end{abstract}

\titlepage

%\newpage

\section{Introduction}

The idea of a hidden sector, that is, a sector that is singlet under the
symmetries of the standard model and interacts with the rest of the
particles only through the Higgs boson (in addition to gravity), sometimes
called ``Higgs portal", was probably first introduced by Veltman and
Yndur\'ain \cite{VY} with the purpose of parametrizing the limit of large
Higgs mass in the standard model.

A more recent motivation of a hidden sector comes from new viable dark
matter models. The existence of dark matter inferred from several different
observations signals the incompleteness of the standard model of the
electroweak interactions \cite{DMReview1}. Perhaps the best motivated
candidate for dark matter are neutralinos arising from supersymmetric
extensions of the standard model but there are other contenders from
different extensions with varying degrees of theoretical motivations 
\cite{DMReview2}.

The simplest extension of the standard model with a natural dark matter
candidate is the addition of a scalar singlet \cite{scalar}. A more recent
role of singlet scalars in dark matter phenomenology appears when one tries
to explain the excess of positrons and electrons seen in some experiments 
as an effect of dark matter annihilation \cite{DMReview2}. In this case, the
annihilation cross section must be a factor of $\mathcal{O} (10^3)$ larger
than the usual cross section determined by the observed dark matter
abundance in the case of thermal relics. The existence of new light singlet scalars
coupling to the dark matter candidate can generate an enhancement
in the annihilation cross section of dark matter particles, known as the
Sommerfeld enhancement \cite{enhancement}. Several studies have been performed
considering this possibility \cite{lightscalars}. In particular, the
coupling of this scalar to the Higgs boson can lead to sizeable effects in
the direct detection of dark matter, as recently emphasized in \cite{direct}.

In this letter we want to study a novel effect of the coupling of the Higgs
boson with another scalar field: the enhancement of the double Higgs
production at the LHC near threshold due to the formation of Higgs-Higgs bound states.

Double Higgs production is an important process to test the structure of the
Higgs boson potential and possible new physics beyond the Standard Model \cite{DoubleHiggs}.
A possible enhancement effect may be of importance since the gluon distribution functions
are largest near the threshold region. 
This is akin to the well known enhancement of top quark pair
production near threshold due to the gluon contribution \cite{threshold}.
Contribution of bound states to the production of supersymmetric particles at the LHC
has also been recently discussed in \cite{kats}. 
The enhancement of the double Higgs production may have observable effects
and therefore can in principle probe the hidden sector.

\section{Enhancement}

The formation of a nonrelativistic Higgs-Higgs bound state, sometimes
referred to as Higgsonium \cite{Grifols} or Higgsium \cite{GristeinTrott} would result in an
increase of the double Higgs production cross section near threshold. 
The possibility of formation of Higgsium can arise within the standard model due
to the Higgs boson self-interactions, but only in the case of a very heavy
Higgs, actually above the unitarity bound \cite{HHbound}. 
Since we will be
interested in an intermediate Higgs boson mass, we will only consider the
contribution of the interaction of the Higgs boson with the hidden sector to this
process.

Double Higgs production within the standard model via the dominant gluon fusion process 
at hadron colliders was
computed by Plehn et al. \cite{PSZ}. They use a form-factor for an effective 
$gghh$ coupling that we will denote by $\Gamma^{(0)}$. This effective
coupling depends dominantly on the total momentum entering the vertex.

We add to the standard model lagrangian a new singlet real scalar $\phi $
with a coupling to the Higgs boson $h$:
\begin{equation}
\mathcal{L}=\mathcal{L}_{SM}+\frac{1}{2}\left( \partial _{\mu }\phi \right)
^{2}-\frac{1}{2}m_{\phi }^{2}\phi ^{2}+guhh\phi
\label{lagrangian}
\end{equation}
where $m_{\phi }$ is the mass of the new scalar, $g$ is a dimensionless
coupling constant and $u$ is an energy scale from this new sector. 
This lagrangian should be thought of as a toy model used to study 
a general phenomenom.
For the moment we will assume that it is already in its diagonal basis for the 
scalar sector. 

Unfortunately there is no full treatment of bound states in quantum field
theory and some approximations must be used, such as the ladder
approximation to the Bethe-Salpeter equation \cite{BS}. We will follow the
standard framework developed for computing threshold effects due to bound
states \cite{threshold} and use the optical theorem to write:
\begin{equation}
\sigma (gg\rightarrow hh)=\frac{1}{s}\mbox{Im}\mathcal{M}(gg\underset{hh}{\rightarrow 
}gg),
\end{equation}
where $\mathcal{M}(gg\underset{hh}{\rightarrow }gg)$ is the amplitude 
for gluon-gluon scattering through a Higgs boson loop.
Our goal will be to compute this amplitude in
the nonrelativistic limit taking into account the contributions from the
hidden sector.

\begin{figure}[h,t,b]
\begin{centering}
\includegraphics[clip,scale=1.0]{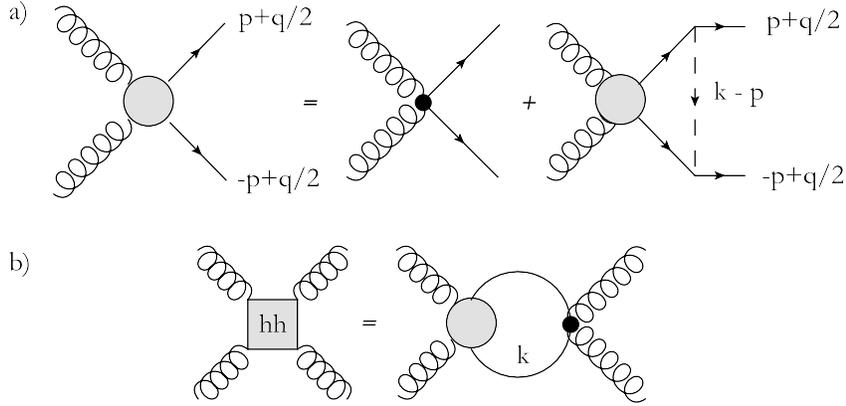}
\caption{\label{fig2} (a) Bethe Salpeter equation for 
${\cal M}(gg\rightarrow hh)$ with hidden sector corrections in the ladder aproach 
(b) Relation between the total scattering amplitude 
${\cal M}(gg\underset{hh}{\rightarrow }gg)$ and $ {\cal M}(gg\rightarrow hh)$.
}
\par\end{centering}
\end{figure}

We first compute the corrections to the effective $gghh$ vertex due to the
hidden scalar exchange by summing all ladder contributions as shown in
Figure (\ref{fig2}a), which results in the integral equation
\begin{eqnarray}
S(q,p)=1 &+&\int \frac{d^{4}k}{\left( 2\pi \right) ^{4}}\;\frac{ig^{2}u^{2}}{%
(p-k)^{2}-m_{\phi }^{2}} \\
&&\frac{i}{(q/2+k)^{2}-m_{h}^{2}-im_{h}\Gamma _{h}}\;\frac{i}{
(q/2-k)^{2}-m_{h}^{2}-im_{h}\Gamma _{h}}\;S(q,k)  \notag
\end{eqnarray}
where $S(q,p)=\Gamma (q,p)/\Gamma ^{(0)}(q,p)$ is the normalized coupling, $%
q $ is the 2-particle bound state momentum and $2p$ is the relative
momentum \footnote{One should note that a including a $hh\phi\phi$ coupling in eq.(\ref{lagrangian}) 
would result in
an additional contribution with a loop of $\phi$ field. This contribution would be suppressed
by the usual $g^2/16 \pi^2$ factor.}. A bound state would appear as a pole at $q^{2}=m_{hh}^{2}$, with 
$m_{hh}<2m_{h}$. Near threshold we write 
$q=(2m_{h}+E,\vec{0})$, where $|E|<<m_{h}$ is
the binding energy for a nonrelativistic bound state. 

In the nonrelativistic limit we keep only the instantaneous part of the $\phi$ propagator. 
Likewise, since we are in the threshold region, we neglect the time dependence of 
the total vertex $gghh$ and for consistency we do the same for the vertex inside the loop in the 
Bethe Salpeter equation (figure \ref{fig2}a).
With these approximations one can perform the $k_{0}$ integral to obtain:
\begin{equation}
S(\vec{q},\vec{p})=1-\frac{g^{2}u^{2}}{8m_{h}^{2}}\int \frac{d^{3}k}{\left(
2\pi \right) ^{3}}\;\frac{1}{(\vec{p}-\vec{k})^{2}-m_{\phi }^{2}}\;\frac{1}{%
\vec{k}^{2}/m_{h}-E-i\Gamma _{h}}\;S(\vec{q},\vec{k})  \label{Green1}
\end{equation}

Setting 
\begin{equation}
G(\vec{q},\vec{p})=\frac{1}{\vec{p}^{2}/m_{h}-E-i\Gamma _{h}}\;S(\vec{q},%
\vec{p})
\end{equation}
and performing a Fourier transform in Eq.(\ref{Green1}) we arrive at a
Schroedinger-like equation, depending of the relative position 
$r$ and center of mass coordinate $r'$ placed at origin:
\begin{equation}
\left[ -\frac{\nabla ^{2}}{m_{h}}-E-i\Gamma _{h}+V(r)\right] G_{V}(\vec{r},\vec{r}^\prime=0
,E)=\delta ^{3}(\vec{r})  \label{Schroedinger}
\end{equation}
with an Yukawa potential given by
\begin{equation}
V(r)=-\frac{g^{2}u^{2}}{8m_{h}^{2}}\frac{e^{-m_{\phi }r}}{r}.
\end{equation}

%%%%%%%%%%%%%%%%%%%%%%%
%%%%%%%%%%%%%%%%%%%%%%%%%%%%%%%%%%%%%%%

Now we can find the amplitude $\mathcal{M}(gg\underset{hh}{\rightarrow }gg)$
by solving the integral equation depicted in Figure (\ref{fig2}b):
\begin{eqnarray}
\mathcal{M}(gg\underset{hh}{\rightarrow }gg) &=&\Gamma ^{(0)}(E) \\
&&\int \frac{d^{4}k}{\left( 2\pi \right) ^{4}}\;\frac{i}{%
(q/2+k)^{2}-m_{h}^{2}-im_{h}\Gamma _{h}}\;\frac{i}{%
(q/2-k)^{2}-m_{h}^{2}-im_{h}\Gamma _{h}}\;\Gamma (q,k)  \notag
\end{eqnarray}

Following similar steps as in the previous calculation we
obtain
\begin{equation}
\mathcal{M}\left( gg\underset{hh}{\rightarrow }gg\right) =\left( \Gamma
^{(0)}(E)\right) ^{2}G_{V}(0,E).
\end{equation}
Therefore the cross section with the contribution from the hidden sector
is
\begin{equation}
\sigma(gg\rightarrow hh)=\sigma_{0}(gg\rightarrow hh)R(E)
\label{cross}
\end{equation}
with the enhancement factor $R(E)$ is given by
\begin{equation}
R(E)=\frac{\mbox{Im}G_{V}(\vec{0},E)}{\mbox{Im}G_{0}(\vec{0},E)}
\end{equation}
where $E=\sqrt{s}-2m_{h}$ is the center-of-mass energy of the Higgsium
from threshold and $G_{0}$ is the solution of the same Schroedinger equation 
(\ref{Schroedinger}) in the absence of a potential.

\section{Finding the enhancement factor}

The solution to eq.(\ref{Schroedinger}) can be obtained following the standard
procedure of defining two independent
solutions $v_1$ and $v_2$ of the corresponding radial homogeneous equation
in the relevant case of zero angular momentum, 
one regular at the origin and the other regular at infinity:
\begin{equation}
G\left( r,r^{\prime },E+i\Gamma _{h}\right) =\left\{ 
\begin{array}{c}
-\frac{m_{h}}{4\pi }\frac{v_{1}\left( r\right) v_{2}\left( r^{\prime
}\right) }{rr^{\prime }}\ \ \ \ \mbox{for} \ 0<r<r^{\prime }<\infty \\ 
-\frac{m_{h}}{4\pi }\frac{v_{1}\left( r^{\prime }\right) v_{2}\left(
r\right) }{rr^{\prime }}\ \ \ \ \mbox{for}\ 0<r^{\prime }<r<\infty \; ,
\end{array}
\right.
\end{equation}
where $v_{1,2}(r)$ is a solution of the equation:
\begin{equation}
\left[ \frac{d^2}{dr^2} +  m_h \left( E+i\Gamma _{h} - V(r) \right) \right] v(r) = 0
\end{equation}

We use the method described in Strassler and Peskin \cite{threshold} to
numerically solve the equation above with the appropriate boundary conditions.
We will use the functions $v_{a}$ and $v_{b}$ with boundary conditions
$v_{a}\left( r\rightarrow 0\right) =r$ and $v_{b}\left( r\rightarrow 0\right) =1$
to write 
\begin{eqnarray}
v_{1}(r) &=& v_{a}(r) \\ 
v_{2}( r) &=& v_{b}(r) +B v_{a}( r).
\end{eqnarray}
Hence $B=\underset{r\rightarrow \infty }{
\lim }\left( -\frac{v_{b}\left( r\right) }{v_{a}\left( r\right) }\right) $
and we can formally calculate the imaginary part of the Green function as 
\footnote{Note that the real part of this limit diverges.}:
\begin{equation}
\mbox{Im} G_{V}( 0,0,E+i\Gamma _{h}) =- \frac{m_h}{4 \pi} \mbox{Im} B
\end{equation}

In our numerical solution the boundary conditions for the Green function is of course 
calculated at a finite value of $r$, and it is very important to keep the Higgs boson width
that guarantees its exponential decay.
%The uncorrected cross section $\sigma_{0}$ and the denominator of the enhancement factor
%in eq.(\ref{cross}) are calculated on-mass-shell.
%\footnote{The dynamical correction is of the order of $\frac{m_h\Gamma _{h}}{\sqrt{s}\Gamma _{h}\left( s\right) }$ %where $s$ is the reaction invariant energy \cite{yokoya}.}

In the absence of a potential it is easy to find that
\begin{equation} G_{0}\left( r,r^{\prime },E+i\Gamma _{h}\right) =
 -\frac{m_{h}}{4\pi }\frac{\sin \left( \lambda r\right) }{\lambda r}\frac{e^{i\lambda r}}{r} 
\end{equation} 
where $\lambda =\sqrt{m(E+i\Gamma _{h})}$ and therefore
\begin{equation} 
\mbox{Im} G_{0}\left( 0,0,E+i\Gamma _{h}\right) =-\frac{m_{h}}{4\pi } \mbox{Re}{\lambda}. 
\label{exact}
\end{equation}

\section{Results and Conclusion} 

We use the results of Plehn {\it et al.} \cite{PSZ} to evaluate the cross section 
without hidden sector corrections \footnote{We will use the $\sqrt{s} >> m_t$ limit for
illustration.}, which can be written as:
\begin{equation} 
\sigma _{0}(gg\rightarrow hh)= \frac{1}{16 \pi s} \left\vert \mathcal{M}(gg\rightarrow hh)\right\vert ^{2}
\sqrt{\left( 1-\frac{4m_{h}^{2}}{s}\right) } .
\end{equation} 

Part of the kinematical factor is cancelled by the denominator of the enhancement factor
of eq.(\ref{cross}) resulting in a total cross section given by
\begin{equation} 
\sigma (gg\rightarrow hh)=\frac{1}{16 \pi s} \left\vert \mathcal{M}(gg\rightarrow hh)\right\vert ^{2}
\sqrt{\left( 1+\frac{2m_{h}}{\sqrt{s}}\right) }
 \; \frac{\mbox{Im} B}{\sqrt{m_h \sqrt{s}}} 
\end{equation} 

The enhancement depends basically on two parameters: the strenght of the interaction $\kappa = 
\frac{g^2 u^2}{8 m_h^2}$ and its range determined by the mass $m_\phi$ of the exchanged boson. 
For $m_\phi = 0$ one is back to the well known Coulomb case.  
As an illustration of this effect we will set the Higgs mass in 180 GeV 
(we checked that our results are weakly dependent 
on the Higgs mass in the intermediate range).

%%%%%%%%%%%%%%%%%%%%%%%
\begin{figure}[h,t,b]
%\begin{minipage}[b]{0.5\linewidth}
\begin{centering}
\includegraphics[clip,scale=0.72]{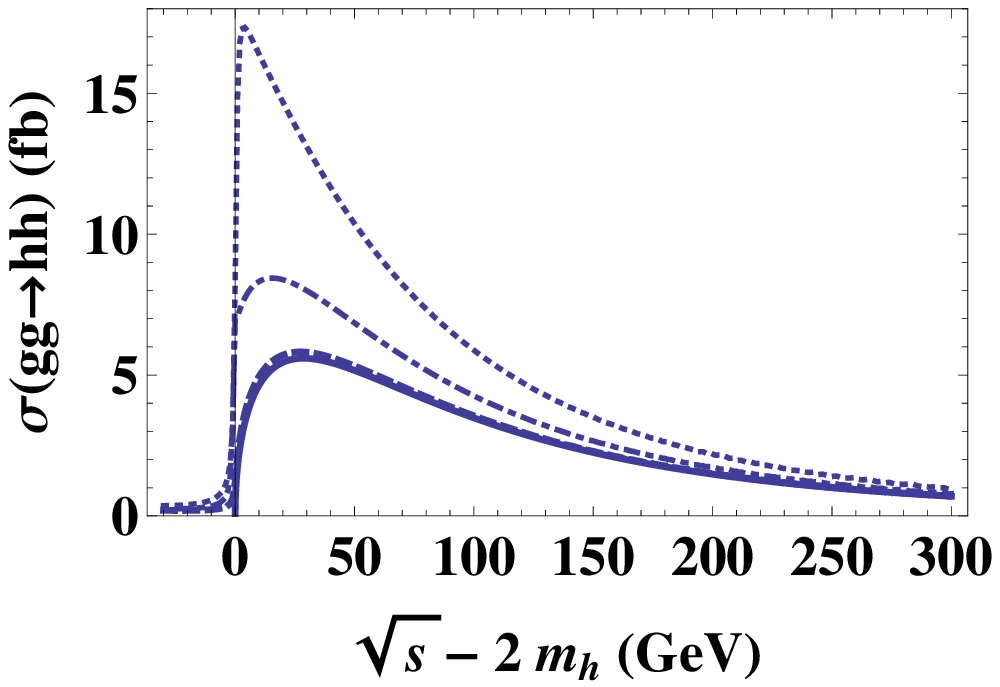}
\includegraphics[clip,scale=0.78]{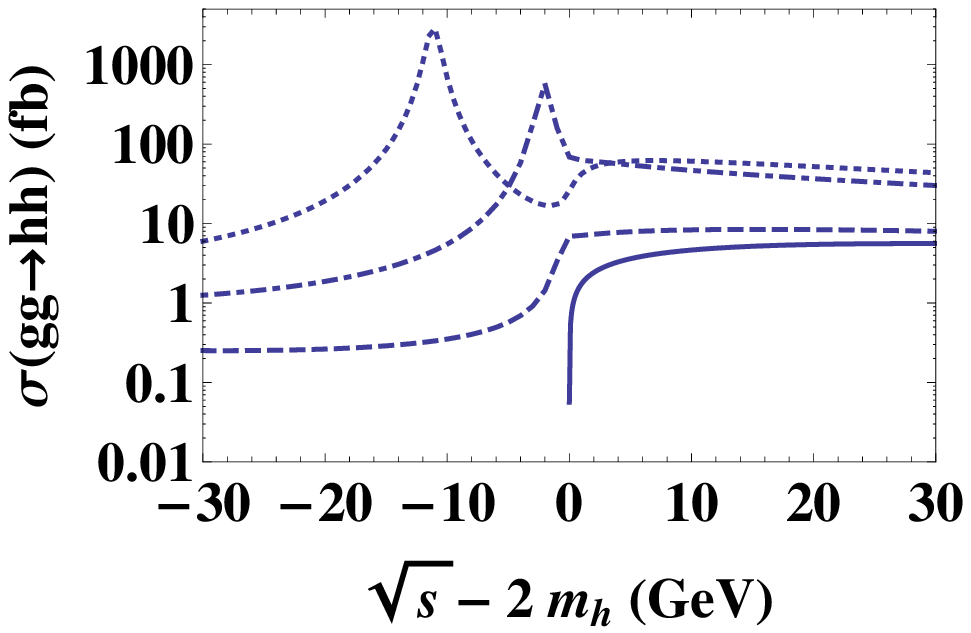}
\caption{\label{fig3} Double higgs production cross section as a function of energy above threshold 
for $m_\phi = 50$ GeV.
Left panel: Double higgs production cross section cross for $\kappa=0.01 ,0.1, \; \mbox{and} \; 0.3$ 
(dashed, dot-dashed and dotted lines, respectively) compared
with the uncorrected result (solid line).
Right panel: Double higgs production cross section near threshold for $\kappa=0.1 ,0.7, \; \mbox{and} \; 1$ 
(dashed, dot-dashed and dotted lines, respectively) compared
with the uncorrected result (solid line). Notice the logarithm scale in this case.
}
\par\end{centering}
%\end{minipage}
%\hspace{2cm}
\end{figure}

%%%%%%%%%%%%%%%%%%%%%%%
%\begin{figure}[h,t,b]
%\begin{minipage}[b]{0.5\linewidth}
%\begin{centering}
%\includegraphics[clip,scale=1.0]{sigmaG.eps}
%\caption{\label{fig3} Cross section for $m_\phi = 50$ GeV and $\kappa=0,0.3, 0.7 \; \mbox{and} \; 1$ 
%(solid lines, bottom to top) compared
%with the uncorrected result (dashed line).
%}
%\par\end{centering}
%\end{minipage}
%\hspace{2cm}
%\end{figure}
%%%%%%%%%%%%%%%%%%%%%%%
%\begin{figure}[h,t,b]
%\begin{centering}
%\includegraphics[clip,scale=1.0]{sigmaM.eps}
%\caption{\label{fig4} Cross section for $\kappa = 0.7$ GeV and $m_\phi=10, 50 \; \mbox{and} \; 100$ GeV 
%(solid lines, top to bottom) compared
%with the uncorrected result (dashed line).
%}
%\par\end{centering}
%\end{figure}
%%%%%%%%%%%%%%%%%%%%%%%%%%%%%%%%%%%%%%%

In Figure (\ref{fig3}) we show the total cross section $\sigma(gg\rightarrow hh)$ for  
different values of the coupling constant for $m_\phi = 50$ GeV. 
For couplings $\kappa < 0.01$ hardly any modification is observed. 
The difference below threshold is due to the finite width effects contained in the function $G$, 
which allows a nonzero cross section below threshold.
We have checked that our numerical code reproduces the analytical result eq.(\ref{exact}) to high accuracy.

The enhancement increases with the
coupling and for values $\kappa > 0.6$ a dramatic difference results from the presence of peaks due to 
Higgs-Higgs bound states.
We find enhancement factors as large
as a factor of ${\cal O} (100)$ in this case, 
which is comparable to what has been found in the study of Sommerfeld 
enhancement for dark matter annihilation cross section in several models \cite{enhancement}.

A couple of comments are in order. First, it is a textbook exercise \cite{Flugge}
to implement a variational method to estimate the energy of the l=0 bound state
in a Yukawa potential\footnote{We thank G.~Krein for pointing this out to us.}. 
In our case the energy is given by:
\begin{equation}
E= -\kappa m_\phi p^3(p-1)/4(p+1)^3
\end{equation}
where p is a solution of the equation $\kappa m_h/m_\phi = (p+1)^3/p(p+3)$.
For $\kappa=1$, $m_h=180$ GeV and $m_\phi=50$ GeV, the values used in our paper,
we obtain $E=-11$ GeV  ($E=-1.8$ GeV for $\kappa=0.7$).
These are in good agreement with our numerical results shown in Figure(\ref{fig3}) and
corroborate that the bound states are nonrelativistic, since the binding energy
is much smaller than the Higgs mass even for these large values of kappa.
Secondly, the nonrelativistic approximation is only valid for a region of $30$ GeV
or so around threshold. The left plot of Figure(\ref{fig3}) is only meant to illustrate
that the corrections are small even at high energies. In our numerical results for the LHC
cross sections below we switched-off the bound state corrections at an energy of $15$ GeV above
threshold.

For lighter $\phi$ masses (${\cal O} (10)$ GeV) the peaks appear for smaller couplings 
at lower energies
and occasionally more than one peak can be seen. For heavier masses, as expected,
bound states are not formed but some enhancement effect is still obtained.  

In order to estimate the corresponding enhancement at the LHC, we have convoluted
the parton level cross section with the gluon distribution function in the proton.
For our estimates we used the Mathematica package of the leading order CTEQ5 \cite{CTEQ}
with factorization and renormalization scales set to $q = 2 m_h$.
In Table (\ref{table}) we present our results for $\sqrt{s}= 7$ and $14$ TeV.
The convolution smooths out the large partonic enhancements but one can still
find cross sections that are as large as 20 times the SM result for extreme values
of the coupling. We find that the enhancement factors are not very sensitive to the 
CM energy, although the absolute values of the production cross section certainly are.

\begin{table}
\begin{center}
\begin{tabular}{|c|c|c|}
\hline
$\kappa$ &  $\sqrt{s} = 7$ TeV  & $\sqrt{s} = 14$ TeV   \\ 
\hline \hline
0.1 & 1.1 & 1.2 \\ \hline
0.3 & 1.5 & 1.6 \\ \hline
0.7 & 5.5 & 6.0 \\ \hline
1.0 & 19 & 21 \\ \hline 
\end{tabular}
\end{center}
\caption{Enhancement factor for double Higgs production cross section at the LHC for different values of the 
coupling $\kappa$ to a hidden sector for $\sqrt{s} = 7$  and $14$ TeV} 
\label{table}
\end{table}

At this point we should note that the simplest model with an additional singlet
would not result in a large enhancement due to the mixing between the scalars, which
requires $\kappa < \frac{\lambda m_\phi^2}{4 m_h^2}$
%The mass eigenstates of the scalars would be given by \cite{wells}:
%\begin{equation}
%m_{1,2}^2 = \left( m_h^2/2 + m_\phi^2/2 \right) \pm \sqrt{ \left( m_h^2/2 - m_\phi^2/2 \right)^2
%+ 4 \kappa m_h^4/2 \lambda },
%\end{equation}
where $\lambda$ is the Higgs boson self-coupling \cite{wells}.
There are a few examples where one could have a strong interaction of the hidden sector with the
Higgs boson, such as considering a more evolved scalar sector or even interactions with hidden
abelian gauge fields or a condensate of fermions.  
We take no position on what the primary motivation might exist for the hidden sector. Our goal is to
consider the general effect of this hidden sector on double Higgs production near threshold.

One may also be concerned with unitarity violation when large
couplings are present. For example, at low energies one gets a s-channel contribution to the
$s$-wave scattering amplitude that goes as $a_0 = g^2 u^2/32 \pi m_\phi^2$; this is less than one
for $g^2 u^2 = {\cal O}(m_h^2)$ and the examples discussed here.

The prospects for detection of such an enhancement require a detailed study that is beyond the 
scope of this short letter. The signal would probably be an excess of 4 gauge boson events near the
double Higgs threshold, where the SM background is small. 
It is amusing to entertain the idea that an investigation 
of the double Higgs cross section near threshold could lead to important information 
about the hidden sector. 

\section*{Acknowledgements} The authors would like to thank  
O.~J.~P.~Eboli, B.~Dobrescu, G.~Krein, F.~Maltoni, M.~Peskin, J.~Rosner and 
H.~Yokoya for valuable discussions. ACAO thanks C.~Grojean for the hospitality at the Cargese
Summer School, where part of this work was done. 
The work of ACAO was supported by a CNPq PhD fellowship and the work of RR was partially supported
by a CNPq research fellowship.

\end{document}